\documentclass[12pt,preprint]{aastex}

\newcommand{\etal}{\it{et al.}\rm}

\slugcomment{To appear in the Astronomical Journal April 2006}
\shorttitle{M33's Variable A}
\shortauthors{Humphreys \etal}

\begin{document}
\title{M33's Variable A  -- A Hypergiant Star More Than 35 Years in Eruption}

\author{Roberta M. Humphreys, Terry J. Jones, Elisha Polomski, Michael Koppelman, Andrew Helton, Kristen McQuinn, Robert D. Gehrz, C. E. Woodward} 
\affil{School of Physics and Astronomy, University of Minnesota,
 Minneapolis, MN 55455}
\email{roberta@aps.umn.edu }

\author{R. Mark Wagner, Karl Gordon, Joannah Hinz}
\affil{Steward Observatory, University of Arizona, Tucson, AZ 85721}

\author{and}

\author{S. P. Willner}
\affil{Harvard-Smithsonian Center for Astrophysics, Cambridge, MA 02138}


\begin{abstract}
Variable A in M33 is a member of a rare class of highly luminous, evolved
stars near the upper luminosity boundary that show sudden and dramatic 
shifts in apparent temperature due to the formation of optically thick 
winds in high mass loss episodes. Recent optical and infrared spectroscopy
and imaging reveal that its ``eruption'' begun in $\sim$1950 has ended, 
{\it lasting $\approx$ 45 yrs}. Our current observations show major changes
in its wind from a cool, dense envelope to a much warmer state surrounded 
by low density gas with rare emission lines of Ca II, [Ca II] and K I. 
Its spectral energy distribution has unexpectedly changed, especially at
the long wavelengths, with a significant decrease in its apparent flux,
while the star remains optically obscured. We conclude that much of its
radiation is now escaping out of our line of sight. We attribute this to the 
changing structure and distribution of its circumstellar ejecta corresponding
to the altered state of its wind as the star recovers from a high mass
loss event. 

\end{abstract} 
\keywords{stars:supergiants, stars:winds,  stars:individual(M33 Var A)}

\section{Introduction}
Variable A in M33 is one of the highly luminous and unstable stars that define the upper luminosity limit
in the Hertzsprung-Russell Diagram for evolved cool stars (see Humphreys and Davidson 1994). It was one 
of the original Hubble -- Sandage variables (Hubble and Sandage 1953) and at its maximum light in 1950
one of the visually brightest stars in M33. Its historical light curve shows a rapid decline from maximum
by more than three magnitudes in less than a year followed by a brief recovery and a second decline.
It had an intermediate F-type spectrum at maximum consistent with its observed colors. After its second decline 
to fainter than 18th magnitude (see light curves in Hubble and Sandage 1953 and Rosino and Bianchini 1973),
no further observations were obtained until our optical and near infrared photometry beginning in 1977.
In 1985-86 Var A had the spectrum of an M-type supergiant (Humphreys, Jones and Gehrz 1987, hereafter HJG), and  
its large infrared excess at 10$\mu$m and spectral energy distribution showed that it was still as luminous 
($5 \times 10^{5} L_{\odot}$) as it was at its maximum light
in the visible. Thus its large photometric and spectral variations had occurred at nearly constant
bolometric luminosity. We concluded that its cool M-type spectrum was produced in a pseudo-photosphere
or optically thick wind formed during a high mass loss episode.

A recent spectrum, twenty years later, reveals another dramatic change. Its spectrum is now that of 
a much warmer star consistent with the warmer photosphere and colors of 50 years ago. In this paper 
we discuss its remarkable spectral changes reminiscent of the variability of $\rho$ Cas (Lobel et al. 2003), 
but on much longer time scales. 
Variable A was  an obvious target for spectroscopy and imaging with the  Spitzer Space Telescope. 
Its resulting spectral energy distribution (.4{\micron} to 24{\micron}) also shows unexpected changes, 
especially at the long wavelengths, most likely corresponding to the changes in its wind and its
impact on its circumstellar medium. In this 
paper we combine all of the available spectroscopy, multi-wavelength photometry and imaging 
to reconstruct the changes in its wind and circumstellar material over the past twenty years. 
In the next section we present our
multi-wavelength observations. The current spectrum and its light curve and spectral energy 
distribution are
described in \S 3 and \S 4.  In \S 5 we discuss Var A's ``eruption'', its duration, and the changing structure of its wind and circumstellar nebula, and in the last section we review its   
relationship to other cool hypergiants and the possible origins of their instability.

\section{Observations and Data Reduction }

Our new observations include blue-red groundbased spectra, polarimetry, and  
spacebased near- and mid-infrared photometry and  spectroscopy. 

\subsection{Spectroscopy with the MMT}

Spectra of Variable A were obtained in 2003 and 2004 with the refurbished MMT now with a 
single 6.5 -- meter mirror. The same dual channel spectrograph was used as in 1985 but with a 
long-slit 1$\farcs$0 wide. Spectra were observed in both the blue and red channels with different gratings resulting in a range of wavelength coverage and spectral resolution. 
The journal of observations is in Table 1. 
The spectra were all reduced in the standard way in IRAF\footnote{IRAF
is written and supported by the IRAF programming group at the National
Optical Astronomy Observatories (NOAO) in Tucson, Arizona. NOAO is operated by
the  Association of Universities for Research in Astronomy (AURA), Inc. under
cooperative agreement with the National Science Foundation} and were flat-fielded, sky subtracted and 
flux and wavelength calibrated. The S/N of the final spectra is wavelength dependent but is 
typically $\sim$ 30 in the continuum and 80 to 90 for the strong H$\alpha$ and H$\beta$ emission
lines.

\subsection{Polarimetry}

Polarimetry at visual wavelengths was performed on 2004 September 27 using OptiPol at the Mount Lemmon Observing Facility 1.5m telescope. OptiPol is a CCD imaging polarimeter using a Wollaston prism in combination with a rotatable achromatic half-waveplate. Images in perpendicular polarizations are split by the Wollaston prism and placed on opposite halves of the CCD. Two polarization standards, HD 204827 (polarized) and HD 212322 (unpolarized),
(Schmidt et al. 1992) were observed.  Due to the faintness of Var A in the visual, we were only able to obtain an upper limit of $P\leq15\% (3 \sigma)$ in the R and I filters (combined). Interstellar polarization could be present, but would very likely be much less than our upper limit.

Photometry was extracted from the polarization data by combining the signal at 
all waveplate positions for each filter. The resulting photometry was placed on 
the Cousins-Kron system by observing Landolt standards (Landolt 1983) in the same 
manner as the polarization standards. The VRI magnitudes are included in Table 2.

\subsection{Spitzer Observations} 

Observations of Variable A were made
using all four bands of the Infrared Array Camera (IRAC) on 2004 January 9
and were repeated on 2004 July 22, August 16 and 2005 January 1 yielding
four epochs of data as part of GTO Program ID 5 to map M33 (P.I Gehrz).
The IRAC instrument (Fazio et al. 2004) uses four detectors at 3.6, 4.5,
5.8, and 8.0{\micron}.  All four detector arrays are 256x256 pixels in
size with mean pixel scales of 1.221, 1.213, 1.222, and 1.220 arcsec/pixel
respectively.  The M33 observations used a 3 point cycling $\sim$ 11 pixel
dither for each position, and the integration time was 10.4 secs per
frame.

The raw data were processed and flux calibrated with version 11.4 of the
Spitzer Science Center (SSC) pipeline.  Details of the calibration and raw
data processing are specified in the IRAC Pipeline Description
Document\footnotemark, version 1.0.
\footnotetext{ssc.spitzer.caltech.edu/irac/dh/PDD.pdf}  Post-BCD(Basic Calibrated Data) processing
was carried out using the 2005 May 9 Linux version of the SSC MOPEX\footnote{Mosaicing and Point Source Extraction} 
software (Makovoz, Khan \& Moshir 2005). Five steps of MOPEX were implemented:
Cosmetic Fix, Final Flat Fielding, Background Matching, Outlier Detection,
and Mosaicking.  The photometry was
measured using a modified version of the ATV IDL photometry package. The
field is relatively empty, therefore a source and sky annulus was chosen
that corresponds to the apertures used in the pipeline calibration
procedures so no aperture correction was necessary. A median sky
annulus was used and uncertainties were calculated using the
prescription in Reach et al. (2005).  The resulting mean magnitudes from
the four observations are in Table 2.

Observations with the Infrared Spectrograph (IRS) were made using the short 
(5-15$\mu$m) and long wavelength (14-28$\mu$m) low resolution 
modules on 2004 August 31 with slit widths of 3$\farcs7$ and 10$\farcs7$, 
respectively. The observations consisted
of 3 cycles of 60 second ramps (effective exposure time) and 3 cycles 
of 120 second ramps in the
short and long wavelength modules respectively. The raw data were
processed and flux calibrated with version 12.0 of the 
SSC pipeline.  Details of the calibration and raw data processing
are specified in the IRS Pipeline Description Document\footnotemark,
version 1.0. \footnotetext{ssc.spitzer.caltech.edu/irs/dh/irsPDDmar30.pdf}
Spectra were extracted from the BCD frames using the Spitzer IRS Custom
Extractor (SPICE) V1.1B16. Individual spectra were then corrected for bad
pixels scaled to the mean, averaged together, and saved in a SPICE-like
format. Uncertainties were estimated from the standard deviation (sample
variance).

Images of M33 at  24, 70, and 160$\mu$m were obtained with the Multiband Imaging Photometer for Spitzer
(MIPS) on 2003 December 29, 2005 February 03, and 2005 September 05.  A
 region approximately $1\arcdeg \times 1\arcdeg$ centered on M33 was
 covered with 4 scan maps.  Each map consisted of medium
 rate scan legs with cross scan offsets of 148\arcsec\ between each
 leg.  The length of the scan legs varied from $0\fdg 75$ to $1\fdg 0$. 
The raw data were processed and flux calibrated with version 3.02  of the
University of Arizona MIPS instrument team Data Analysis Tool (Gordon et
al. 2005).  Extra processing steps on each
image were applied before mosaicking using programs written
specifically to improve the reductions in large, well-resolved galaxies.
The first epoch 70{\micron} observations (29 Dec 2003)
were taken before the operating parameters of this array were finalized, 
and therefore  suffer from significantly larger detector transients than
the second epoch data.  All three  epochs  were used
for the final mosaics at 24 and 160~\micron, but only the 2nd and 3rd epoch data
were used for the final 70~\micron\ mosaic. The mean magnitude at 24{\micron} is
included in Table 2; however, Var A was not detected at either 70
or 160{\micron}, so only  upper limits are given for those wavelengths.

\section{The Spectrum}

The spectrum of Variable A observed in 1985 and 1986 (HJG) was 
that of an M-type star with strong TiO bands and a weak H$\alpha$ emission line. 
This spectrum was somewhat of a surprise at that time because the other Hubble-Sandage
variables in M31 and M33 were hot stars (Humphreys 1975, 1978) and belonged to the group of unstable massive
stars we now call Luminous Blue Variables (see Humphreys and Davidson 1994 for a review).
Assuming that Var A's normal state was that of an F-type supergiant, we attributed this
M-type spectrum to the formation of a cool optically thick wind corresponding to its
rapid decline in apparent brightness and color shift 30 years earlier. 

Our recent spectra from 2003 and 2004 reveal another dramatic shift (see Figure 1). 
The TiO bands are gone. The spectrum has returned to  a much
warmer apparent temperature with  strong hydrogen emission, absorption lines 
appropriate for  an early  F type to an early G-type star, depending on which lines are used,  plus unusual  emission lines including K I ($\lambda\lambda$7665,7699),
the near-infrared Ca II triplet ($\lambda\lambda$ 8498,8542) and [Ca II] at $\lambda$7292 and $\lambda$7324 \footnote{The Na I 
'D' lines may also be in emission, although measurements in that spectral region are complicated
by subtraction of the strong night sky lines.} 
Thus the star has returned to its previous warmer state; although, at 
this time it is not possible to know if the absorption lines are representative of the star's
true photosphere after the dissipation of its  
cool, dense cirumstellar envelope or are formed in a warm wind which is consistent 
with the range of identified lines. Var A's recent optical colors (Table 2) are consistent
with its much warmer spectrum; however, although it is no longer red, Var A has 
remained faint. 

The strong emission lines and representative absorption lines are listed in Table 3 with 
their Heliocentric velocities and equivalent widths.  Because of the low resolution many 
of the absorption lines are blended. The expected Heliocentric velocity of Var A at 
its distance from the center of M33 would be about -130 km s$^{-1}$ from the H I radial velocity map (Plate 6) in Newton (1980). With respect to this expected velocity, most of the emission and absorption lines are blue shifted  which could be due to expansion of 
the ejecta or to the star's systemic motion, while the K I emission is redshifted. 
There are also several unidentified emission and absorption lines, and some of the stronger 
ones are listed in Table 4. 

Var A  still has  an extensive but low-density circumstellar envelope
responsible for its strong hydrogen and peculiar emission. Both the H$\alpha$ and H$\beta$ emission lines show prominent wings extending to $\pm$ 800 --
900 km s$^{-1}$ (see Figure 2) presumably due to Thomson scattering. There is no obvious
associated P Cygni absorption; although, at this resolution it may be  difficult to tell.
Emission lines of K I, the Ca II infrared triplet and [Ca II] are rare in astronomical 
spectra, but are observed in
some of the other ``cool hypergiants''  with extensive circumstellar material such 
as IRC+10420 
(Ca II, [Ca II], K I; Jones et al 1993, Humphreys et al 2002) and VY CMa (K I; Wallerstein 1958 and numerous subsequent papers).
K I emission is only weakly present in the ejecta of IRC +10420 and also in $\rho$ Cas (Lobel 1997).
However, in VY CMa it is not only the strongest emission line, but is also stronger there than in any other known late-type star with peak fluxes several times the continuum. In Var A, the K I lines are second
only to H$\alpha$ and H$\beta$ in relative strength with peak fluxes about 1.5 times the continuum 
level. Their doublet ratio also indicates that the lines  are optically thick as in VY CMa, and
also like VY CMa, the $\lambda$7699 line is the stronger. Figure 3 shows the region around the K I lines.  

K I emission is  normally
attributed to resonance scattering. However, averaging over directions in a simple model, resonance scattering neither augments nor diminishes the net emergent flux; apparent ``emission'' must be balanced by an equal amount of ``absorption'', possibly redshifted or blueshifted or visible along a different line of sight. In a recent paper on VY CMa,  Humphreys {\etal} 2005 (see the Appendix and their Figure 14)  suggested a model wherein the K I ``emission'' component dominates in the emergent spectrum from multiple scatterings from large, dusty 
condensations in a roughly spherical inhomogeneous shell. In that model, lines of sight where resonance scattering would produce net ``absorption'' are preferentially hidden  because they occur in denser localities, while lines of sight with net ``emission'' are more favorable for photon escape. (For a more complete description, see the paper cited above.)  
The K I emission lines in Var A are not as extreme as in VY CMa, but their ratios
indicate that they are produced in similar conditions. The redshift of these lines relative to other features (Table 3) may be difficult to explain in this model, but this depends on geometrical details. Alternatively, the gas and dust around Var A may be non-spherical and we may view it from a direction that favors the ``emission'' component of resonance scattering. As we will see in \S 5 a non-spherical geometry is likely. The probable presence of large, dusty condensations will be relevant to our later discussion in sections 4 and 5. 

Like K I, [Ca II] is rarely observed in emission because the atoms are normally collisionally 
de-excited back to the ground state that produces the Ca II H and K lines. 
The transition that produces the Ca II infrared triplet emission leaves 
the atoms in the upper level for the forbidden lines, some of which must be radiatively de-excited to produce the forbidden emission.   Strong emission from the Ca II infrared triplet 
and [Ca II] emission are also  seen in the spectrum of IRC+10420.  Therefore, in 
two of our 
previous papers on IRC+10420 (Jones et al 1993, Humphreys et al 2002), 
we estimated the relative number of de-excited photons from the ratios of their combined equivalent
widths ([Ca II] to Ca II\footnote{The Ca II $\lambda$8662 line was not covered by our spectra. Therefore, we estimated its equivalent width from the expected line ratios of the Ca II triplet compared with the measurements of the other two lines.}) to their corresponding continuum fluxes. Following this same procedure
for Var A, we find that $\approx$ 27\% of the photons produce the [Ca II], and $n_{e}$ is only
2.7 times the critical density for  radiative de-excitation. This is very similar to our
results for IRC+10420, and shows that both stars possess extended regions of very low density gas beyond their dense winds or photospheres.

The IRS spectra from 5 to 15{\micron} and 14 to 28{\micron} show
that  Var A also has prominent silicate emission at 9.7{\micron} and 18{\micron}. The
9.7{\micron} feature is quite strong in emission, at least as strong as the
9.7{\micron} emission feature in the Galactic hypergiant IRC+10420 (Jones {\etal} 1993),
see Figures 5 and 6.  The ratio of the 9.7 to 18{\micron} emission features 
 is typical for optically thin dust shells in Galactic
red giants and supergiants (e.g. Little-Marenin \& Little 1990).

\section{The Light Curve and Spectral Energy Distribution}

Figure 4 shows Var A's light curve from 1950 to the present   
based on the work by  Hubble and Sandage (1953)
and  Rosino and Bianchini (1973) together with the photoelectic and 
CCD blue and visual photometry between 1977 and 1986 from HJG,
photographic B photometry from 1982 to 1990 (Kurtev {\etal} 1999),  and
CCD photometry from 2000-01  (Massey 2005).  The older photographic
color index has been converted to an approximate $\bv$ color (Allen 1963).
The most recent data for 2004 are derived from integration of our flux calibrated spectra
over the standard band passes for B, V, and R magnitudes,
and from the VRI frames used for the polarimetry observations. These  two independent  
methods give  consistent results.
Hubble and Sandage (1953) originally suggested that the decline of $\sim$ 3 to 3.5 magnitudes 
in the photographic
blue was consistent with a shift in bolometric correction 
and the corresponding  change in color index. The 1985-86  
spectrum confirmed that the star's apparent 
energy distribution had indeed shifted to a much cooler temperature. 
However, between 1954 and 1986 the star  faded an 
additional two magnitudes in the blue which we attribute to 
extinction by circumstellar  dust. 

In Table 2 we summarize the multi-wavelength photometry for Var A from 
HJG, an unpublished JHK measurement from 1992 by
Gehrz and C. Woodward, optical photometry from Massey(2005) and more recent photometry from the calibrated optical 
spectrum and polarimetry, JHK photometry from the 2MASS All--Sky Point Source Catalog, and the  
mean 3.6 to 8.5{\micron} IRAC measurements from Spitzer. The current 24{\micron} point is from the  Spitzer MIPS observation.  
Figure 5 shows Var A's present broadband  spectral energy distribution together with the IRS  
spectrum compared with its energy distribution from 1986.  This comparison reveals several very significant changes.

The most oustanding difference is the large drop in flux by a factor of three at 10{\micron}, 
and by $\sim$ 5 times at 18{\micron}.  
Note that in 1986 our measured 10{\micron} flux showed that the star's total energy 
output had remained constant and that it was still as luminous as at its maximum light
in 1950. The recent decline in the long wavelength flux has important implications for possible
dissipation and/or destruction of the reradiating dust and for any model for the star's
changing wind and distribution of the circumstellar material. Consequently we have 
carefully checked
the earlier mid-infrared photometry and the IRAS data used by HJG, and are confident in the earlier, 
higher fluxes for the following reasons.  

There is clearly a source in the IRAS $12~\mu m$ coadded maps at the position
 of Var A. A subsequent analysis of IRAS observations of M33 by Rice et al. (1990) reached 
 a similar conclusion. Rice et al. derived a $12~\mu m$ flux (.10 Jy) and an upper limit for 
 the $25~\mu m$ flux (.09 Jy) consistent with the values we reported in HJG. 
 There is also no evidence in the current Spitzer observations of another source, such
  as a bright HII region, that could have contributed more than 25\% to the IRAS fluxes.

The ground based  10{\micron} magnitude used in HJG (see Table 2 this paper) was 
derived from  data 
from four separate observing runs. These observing runs on the IRTF and WIRO 
telescopes resulted in one $4\sigma$ detection and three $3\sigma$ upper 
   limits. Combining these observations, HJG computed a $4\sigma$ value for the 10{\micron} 
   flux of Var A which  is completely independent of, but entirely consistent with, the IRAS $12~\mu m$ flux.

Finally, we note that the luminosity of Var A in the mid-infrared reported in
HJG was very close to the pre-outburst optical luminosity of the star. Observations 
of yellow and red hypergiants as well as the Luminous Blue Variables
show that these stars maintain  essentially constant
luminosity as they undergo variations in their apparent temperature 
(see Humphreys and Davidson 1994). 
If dust formed in the material shed by Var A and completely covered the star, then it would 
absorb all of the stellar luminosity and reradiate it in the infrared, as observed by 
HJG. The present decrease in the 10{\micron} and 18{\micron} flux compared with Var A's 
luminosity at maximum light means that a significant fraction of the star's light must
be escaping elsewhere. The lack of measurable flux at even longer wavelengths (70 and 160$\micron$) rules out the presence of cooler dust that could be re-radiating the missing flux. Furthermore
there has not been sufficient time for the circumstellar material to have expanded to those 
much larger distances from the star.

The fading of its near-IR flux (from 2MASS) very likely corresponds to
a combination of the shift in the energy distribution and the possible dissipation of 
warm dust nearer the star due to changes in the wind as its  density decreased and 
radiation from warmer layers escaped. The optical colors between 2000 and 2004 show the object
getting slightly bluer which may be due to a continued decrease in wind density.  The recent photometry also shows that
the return to a warmer apparent photosphere was not accompanied by
a visual brightening. The most likely explanation is that its present faintness 
in the optical is due to obscuration 
by circumstellar dust, although it is not red, suggesting that the
circumstellar extinction is relatively neutral due to grains larger than is 
typical for interstellar and circumstellar dust, but still consistent with the silicate
emission. Compared with its visual maximum,
when Var A apparently had a comparable temperature, this implies that it now has
approximately four magnitudes of circumstellar extinction in the visual. 

Variable A's eruption, the changes in its wind and  circumstellar 
material are discussed in the next section.

\section{Discussion -- The Wind and Circumstellar Material in Transition}

Our current observations reveal some remarkable changes in the spectrum and
energy distribution of Variable A. Although, we do not have a more complete
spectroscopic and photometric record over the past 20 years,  our
observations do allow us to put together a picture of a very luminous, unstable
star experiencing  major changes in the structure of its wind and circumstellar material
at the end of a  high mass loss event. 

Variable A  still  had its cool, dense wind and was in a high mass loss phase 35 years
 after it rapid decline in 1951. Thus it was in ``eruption'' for at least that
 long. The photometric record also shows that Var A faded two magnitudes
 between 1954 and 1986 presumably due to the formation of dust, and may have created 
 an additional two magnitudes of circumstellar extinction since then. 
 Our recent spectrum 20 years later shows that the star's F-type photosphere has returned.
 When did this transition occur? The star has remained faint even though the spectrum
 has presumably recovered. The variation in its near-infrared photometry may correspond
 to changes in the wind accompanying the subsidence of its cool, dense wind phase. 
 The 2MASS observations obtained in Dec. 1997 show that
 Var A  had faded significantly in the near-infrared. However, the JHK photometry  
 from 1992 does not show any change from the earlier data in 1986.
Assuming that the near-infrared variation is due to the wind in transition, then  
 Var A's  eruption or dense wind stage may have lasted between 41 and 46 years!  
 
 How long the transition back to a warmer temperature may have taken is
 uncertain, but we note that the initial formation of the dense wind took approximately 
 a year based on its 1950's light curve or at most two years if we include the brief 
 recovery and second decline. Using the record of $\rho$ Cas, which shows similar outbursts
 or shell episodes as an example, the onset and recovery timescales are comparable, and in
 $\rho$ Cas they occur very rapidly, in only a couple of months. 
 Thus many of the changes we discuss below may have occurred over only one to two years
 at most for the spectroscopic changes and perhaps up to five years (1992 to 1997) 
 or so for the onset of the changes in the distribution and structure of the circumstellar 
 material.
 The transition in the wind from an M-type false-photosphere to a warmer F-type star,
 could reasonably have occurred on even shorter timescales. Although we do not have a direct
 measurement of Var A's wind speed, the expansion velocities for the winds and ejecta in IRC+10420
 and VY CMa are 40 -- 60 km s$^{-1}$ and 35 km s$^{-1}$ for the envelope expansion during 
 $\rho$ Cas's recent episiode. Assuming 50 km s$^{-1}$ for the wind speed, Var A's 
 envelope could have made these transitions in as short  as  3/4's of a year, consistent
 with its variations during the early 1950's. 
 
To explain Var A's observed energy distribution in 1986, HJG  proposed a simple model 
of a cool, dense false-photosphere with an obscuring torus and reflection nebulae at 
the poles. This model combined extinction by circumstellar dust with the blueing 
effect of scattering by dust grains and assumed that most of the flux was radiated in the
mid-infrared. Most of the visible light was scattered and not viewed through the 
intervening material. In this model the visible light would very probably have been more
highly polarized than the upper limit of 15\% reported here. There was no optical 
polarimetry in 1986, so we cannot check this model or verify that there has been a change.

 Despite its warmer apparent temperature and the corresponding
 change in the blue-visual energy distribution, Var A has not brightened in the visual,
and  comparison with its maximum light implies $\approx$ 4 magnitudes of circumstellar 
 extinction currently in the line of sight.  HJG showed that the visual interstellar extinction for Var A was 
 about 0.6 to 0.8 mag which is typical for stars in M33. With its current $\bv$ color of 0.8 to 0.9 
 this suggests that its current true color is 0.6 to 0.4, appropriate for a late F-type star, 
 and in comparison with its colors at maximum light ($\sim$ .4), implies virtually no 
 circumstellar  reddening   in the visual.  As we mentioned previously, the dust grains 
 must be large enough 
 to provide the neutral extinction in the visual which is also consistent with  the scattering requirements 
 for the K I emission. Our polarimetry upper limit of 15\% suggests that
 the optical light at present is not due mostly to scattered light.  
 If we adopt 30\% for the polarization of a star completely blocked by circumstellar dust
 and visible only in reflected light (Johnson and Jones 1991), then at most half the 
 visible light from Var A is from scattered photospheric radiation. The polarimetry also
rules out any current asymmetries such as bipolar lobes although reflection nebulosity
in a more spherical distribution could still be present.

The most dramatic change in Var A's energy distribution is the apparent decrease in its
total flux. Since it is highly unlikely that the total luminosity of the star would have
declined by more than a magnitude, the energy must  now be 
escaping in some direction other than along our line of sight. Several massive stars
are now known to have asymmetrical winds or bipolar outflows. The most notable is $\eta$
Car with a latitude dependent wind (Smith et al 2002) that is both faster and denser at
the poles, although $\eta$ Car is a very different kind of star. 
However, given the evidence from other evolved, massive stars including the cool hypergiants
like IRC+10420 and VY CMa, for irregularities and density variations in their
circumstellar ejecta, it is likely that Var A's dusty shroud is not uniform or
completely opaque in all directions. 
In the remaining discussion
of  Var A's wind and circumstellar material we will assume that the
missing radiation is escaping through large,  low density regions, even holes,
in the obscuring material, and based on the timescales discussed above for 
the duration of the eruption
and the recovery, we'll assume that Var A has been in this warmer state for approximately
ten years.

HJG fit  Var A's 1986 mid-infrared flux  by a 370$\arcdeg$ BB which implies a dusty
zone or shell with a radius of $\sim$ 400 AU from the star. The energy distribution had  
a significant near-infrared flux due to warmer dust closer to the star, so the dusty 
zone probably extended from 100 to 400 AU or so. Figure 4  shows that Var A slowly 
faded over 30 years due to increased circumstellar obscuration 
presumably as the amount and density of the  
obscuring material increased. HJG had estimated a mass loss rate of $2 \times 10^{-4}$ M$_{\odot}$ yr$^{-1}$ from Elitzur's (1981) formulation  assuming radiation coupling to the grains. But when Var A's optically thick wind ceased and quickly subsided
back to a warmer state, then its mass loss rate would very likely have decreased as well. 
For example,  
the mass loss rates of the LBV's during their eruptions are typically 10 to 100 times
that during their quiescent stage, and  the normal mass loss rates of $\rho$ Cas and 
HR 8752, two hypergiants of comparable luminosity and temperature, are $\sim$ $10^{-5}$ M$_{\odot}$ 
yr $^{-1}$.  In addition to a less dense wind and a lower mass loss rate, Var A may 
also have a higher wind speed now. Wind speeds of 100 to 200 km s$^{-1}$ are typical 
of normal A to F-type supergiants. Following HJG's procedure, using Var A's apparent luminosity inferred from its current 10$\micron$ flux, and assuming that the optical depth is near one in our line 
of sight, we find a current mass loss rate of $\sim$ $6.7  \times 10^{-5}$ M$_{\odot}$ yr$^{-1}$ for 
a 50 km s$^{-1}$ wind. But if the wind speed is higher, the mass loss rate will be 
$\sim$ 2 -- 3 $\times 10^{-5}$ M$_{\odot}$ yr$^{-1}$. 

In the approximately 10 years since its transition,
Var A's lower density and possibly faster wind would have reached 100 to 300 AU, the 
region of the dusty zone. Consequently, the dusty material may not be replenished as 
efficiently as in the previous dense wind, higher mass loss stage. 
So as the dusty zone has continued to expand during this same period, the lower
density regions or gaps will also have enlarged allowing more radiation to escape.
If the dusty distribution is flattened, as HJG suggested, this is most likely to 
occur at the poles. 

Therefore we propose that large dusty condensations in a flattened 
distribution, possibly a torus, currently obscure our direct view of Var A. 
The K I emission lines are produced by resonance scattering in this dusty zone. 
In addition, a very low density gas or wind responsible for  the hydrogen, 
Ca II, and peculiar [Ca II] emission fills the region between the star's photosphere 
and the dusty zone.  More than half  of the radiation is escaping from our line of
sight, presumably from low density regions, all or most of which must
be out of our line of sight. Furthermore, the polarimetry measurements rule out a 
strongly bipolar structure with asymmetrically scattered light. This then suggests a 
unique geometry even if
the gaps are restricted to the polar regions.  The dusty zone must either be 
nearly aligned with our line of sight in such a way to also block most of the 
escaping radiation from being reflected back into our line of sight or else there
is very little reflecting nebulosity.

\section{Discussion -- Variable A and the Cool Hypergiants}  

HJG derived a total luminosity for Var A of $5 \times 10^{5}$ L$_{\odot}$  based on both its
luminosity at its visual maximum (M$_{v}$ $= -9.4$ to $-9.6$ mag)  and its mid-infrared flux in 1986. 
As we've already emphasized the two agreed. Its position on the HR diagram in its two states, cool
dense wind and warm photosphere or wind, can be seen in Figure 
4 in HJG and also on Figure 14 in Humphreys et al (2002) together with other very
luminous evolved cool hypergiants some of which have already been mentioned in this paper. Var A 
is one of several luminous stars in our galaxy and others that define the upper luminosity
boundary for evolved cool stars (Humphreys and Davidson 1979, 1994).

Var A shows spectral and photometric characteristics in common with several of these stars,
especially IRC~+10420 and $\rho$ Cas.
Spectroscopically, it is most like the post--red supergiant IRC~+10420 with its strong hydrogen emission,
and the Ca II and [Ca II] emission lines. IRC+10420 has a large infrared excess, is
a powerful OH maser and has an extended and complex reflection nebula (Humphreys et al 1997).  but it does not have significant circumstellar reddening. Figure 6 shows a comparison
of their spectral energy distributions. Var A and IRC+10420  have comparable
luminosities with strong 9.7{\micron} silicate emission features and
 gently rising continua from 1-8{\micron}. IRC+10420 has high 
interstellar extinction in the visual that causes the rapid drop in flux
at wavelengths shorter than 1{\micron}. Emission from warm dust near the star
  contributes to its excess  radiation longwards of 2{\micron}. This  is  
observed in other evolved stars with extensive circumstellar ejecta, and   
may also be the case for Var A (see HJG). The 18{\micron} silicate feature 
is unusually  strong in IRC+10420 compared to the more ``normal'' feature  in Var A. 
Based on laboratory investigations, Nuth and Hecht (1990) suggest that a relatively stronger 
18{\micron} feature indicates grains that are more highly processed,
implying that the   dust associated with Var A formed relatively recently and is not 
the remnant of an old mass-loss phase in the star's evolution.
We do not have any record of a prior cool, dense wind state in IRC+10420, although 
Humphreys et al (2002) concluded that with
its very high mass loss rate (3 -- 6 $\times 10^{-4} M_{\odot}$ yr$^{-1}$),
IRC+10420's current warm wind ($\approx$ 8000 -- 9000$\arcdeg$) is optically thick. Interestingly,
the light curves of both Var A and IRC+10420 show a long period during which they slowly
increased in apparent brightness (see Hubble and Sandage 1953, Gottleib and Liller 1978).
In Var A this slow rise in brightness culminated in its 1950's ejection episode, while
in IRC+10420, the star has been at essentially constant visual brightness since 1970.
It is interesting to speculate, that both of these stars were recovering from a previous
eruption during which they had suffered significant circumstellar extinction which
had slowly dissipated in the preceding decades.

The other example, $\rho$ Cas, is best known for its historical and recent high mass 
loss episodes during which 
it produces a cool dense wind with the corresponding spectral changes from a warm  
F-type photosphere to an M-type with TiO bands. However, the durations of its episodes
are quite short. The most recent in 2000 lasted only 200 days (Lobel et al 2003) while the previous recorded
1945--6  event lasted 1 to 2 years (Beardsley 1953, Bidelman and McKellar 1957). However, $\rho$ Cas and a similar star HR 8752, have very little
if any circumstellar dust  and neither has any associated visible nebulosity in 
HST/WFPC2 images (Schuster, Humphreys and Marengo 2005). 
The  much longer duration of Var A's optically thick wind and high mass loss stage
may account for its obscuring circumstellar material, although this brings up issues
concerning the sustainability of the ejection and the origin of the underlying instability. 
    
de Jager (1998) has suggested that the intermediate temperature hypergiants are post--red 
supergiants  which in their evolution to warmer temperatures enter a temperature range
(6000 -- 9000 K) with increased dynamical instability where high mass loss episodes occur. 
Due to increased mass loss in the red supergiant  stage the most massive, most luminous 
of these stars, near the upper luminosity boundary, may lose enough mass to bring 
them close to the classical
Eddington limit ($(L/M)_{Edd} \; = \; 4 {\pi} c G / {\kappa}$). Then in their post--red
supergiant evolution,
as the  apparent temperature increases above 6500 K, the ionization and opacity increase
rapidly, and  the ``modified Eddington limit''\footnote{$\kappa$ is temperature dependent.}
becomes important (see Humphreys and Davidson 1994, Humphreys et al. 2002). When combined 
with other atmospheric effects such as ionization of hydrogen
and pulsation, this leads to high mass loss episodes and very strong winds. Indeed, proximity to
the modified Eddington limit may amplify  other instabilities making their effects more extreme. 
IRC~+10420, $\rho$ Cas and HR8752 are most often cited as examples of evolved stars in this
stage, and Var A undoubtedly belongs to this group. 

$\rho$ Cas's recent  high mass loss episode was preceded by a period of pulsational
instability and Lobel et al (2003) demonstrated that the outburst was initated by the 
release of ionization energy due the recombination of hydrogen as the atmosphere cooled
during the expansion. This mechanism explains the short episodes  in $\rho$ Cas, but is
it adequate to maintain the long term optically thick wind observed in Var A?    
Perhaps a deeper instability, triggered  by an atmospheric phonomenon, is responsible
for maintaining the ``eruption'' for more than 35 years. 

Variable A's high mass loss stage has apparently ended.
If in 1950 we had our current ground and spacebased
optical and infrared instruments, we could have observed the changing characteristics
of its  ejecta, the formation and dissipation of dust, over a 50 year period. 
Nevertheless, despite its present apparent faintness,
continued observation of Variable A in the optical and near-infrared is important to 
monitor further changes. It is the first object of this type for 
which we have found significant changes in its energy distribution corresponding 
to a recent high mass loss event. Variable A thus  presents us with the opportunity to  continue to
observe the changing structure of its circumstellar ejecta as it recovers from a long term  ``eruption''.

\acknowledgments
It is a pleasure to thank Kris Davidson for useful discussions about winds and instabilities in
evolved massive stars. We are especially grateful to Phil Massey for providing the UBVRI photometry 
of Var A in advance of publication from  the Local Group Galaxies Survey (Massey
{\etal} 2002). This work was supported in part by the University of Minnesota.
It is based on observations made with the Spitzer Space 
Telescope, which is operated by the Jet Propulsion Laboratory, California 
Institute of Technology, under NASA contract 1407. Support for the authors was
provided by NASA through contracts 1256406 and 1215746 issued by
JPL/Caltech.

{} 

\clearpage 

\begin{figure}
\epsscale{1.0}
\plotone{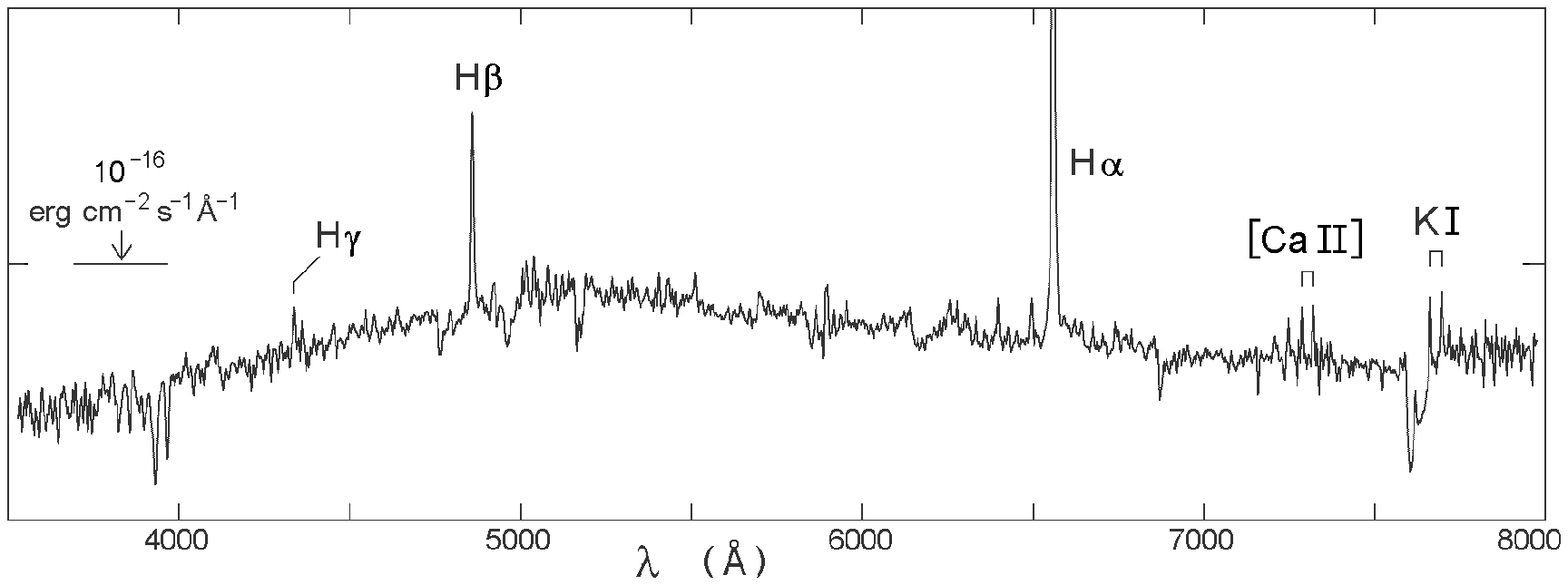}
\caption{The optical spectrum from $\sim$ 3600{\AA} to 8000{\AA}. The strongest emission lines are marked. The lower boundary represents zero flux, and a level of
  $10^{-16}$ erg cm$^{-2}$ s$^{-1}$ {\AA}$^{-1}$ is marked
    near the left side.}  
\label{Spectrum}
\end{figure}

\clearpage

\begin{figure}
\epsscale{0.75}
\plotone{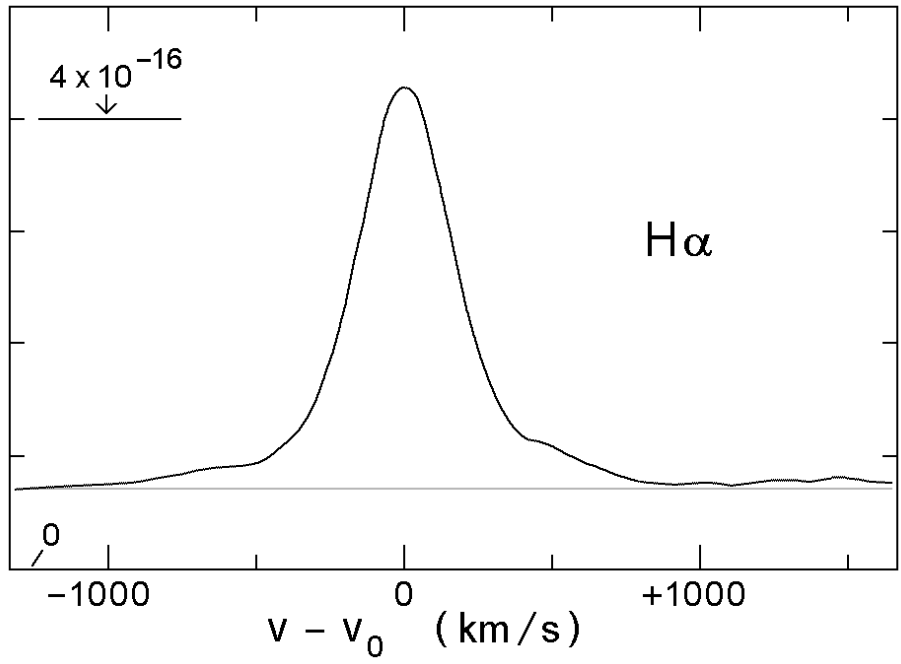}
\caption{The H$\alpha$ profile showing the broad Thomson scattering wings. The x-axis is relative to the
centroid of the profile. H$\alpha$ has a heliocentric velocity of -150 km s$^{-1}$.
The lower boundary represents zero flux, and a level of
  $4 {\times} 10^{-16}$ erg cm$^{-2}$ s$^{-1}$ {\AA}$^{-1}$ is marked in the 
  upper left corner.}    
\label{Halpha}
\end{figure}

\clearpage

\begin{figure}
\epsscale{0.75}
\plotone{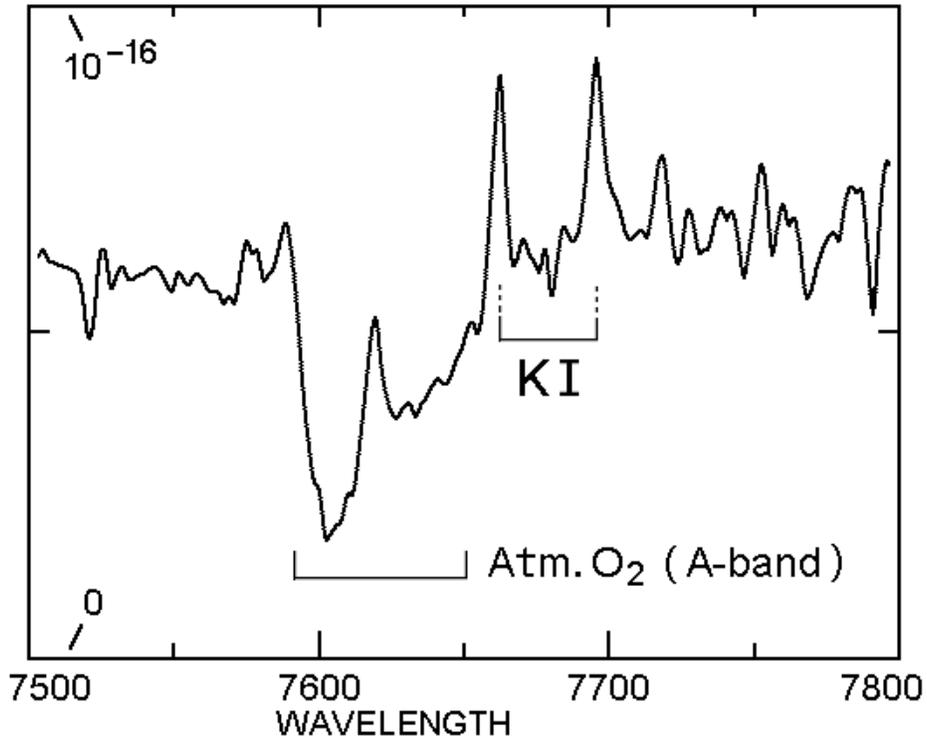}
\caption{The far--red spectrum showing the region of the K~I emission lines and the atmospheric A--band. The bottom and top boundaries represent flux levels of
  zero and $10^{-16}$ erg cm$^{-2}$ s$^{-1}$ {\AA}$^{-1}$ respectively. }
\label{KI}
\end{figure}

\clearpage

\begin{figure}
\epsscale{0.66}
\plotone{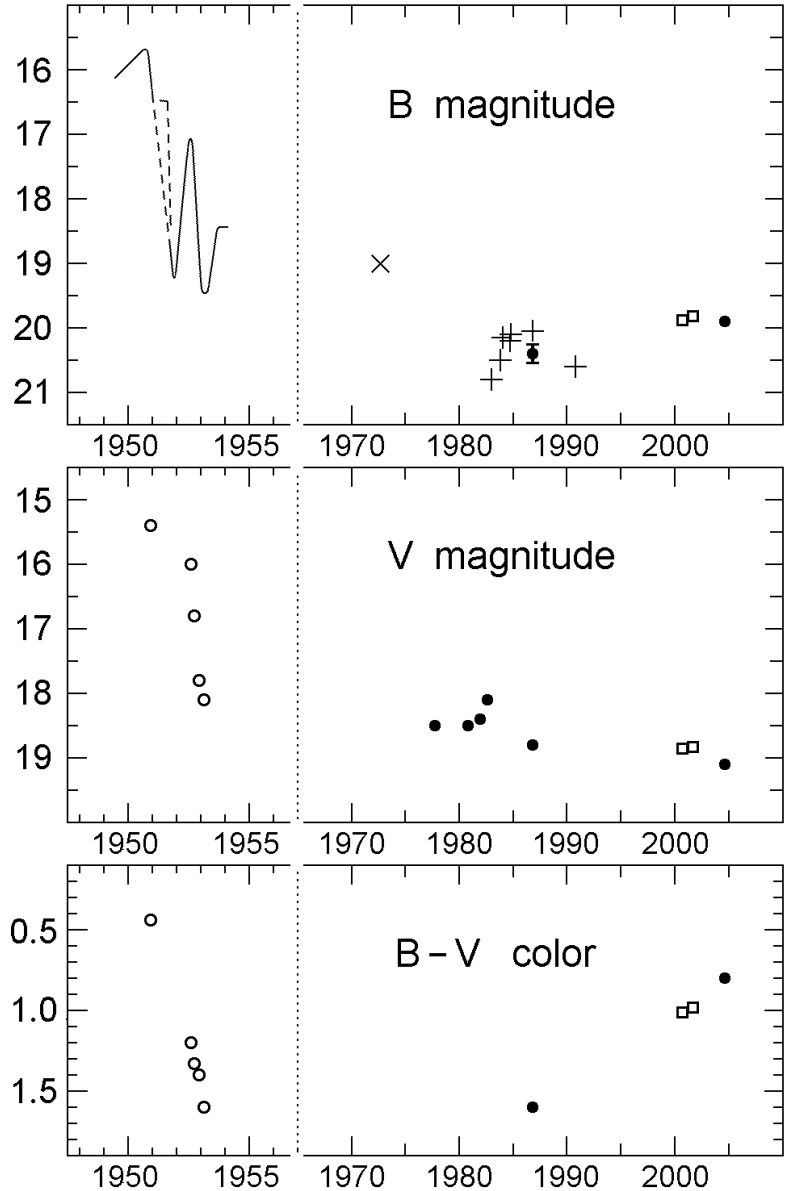}
\caption{Variable A's light curve from 1950 to the present. The upper panel shows the photographic and 
B band magnitudes. The solid and dashed lines are from Hubble \& Sandage (1953) and Rosino \& Bianchini 
(1973). The middle and lower panels show the variability in the V band and the {\bv} color. The different symbols are from Hubble \& Sandage (1953) {\it open circles}, Rosino \& Bianchini (1973) {\it tipped crosses}, Kurtev {\etal} (1999) {\it crosses}, Massey (2005) {\it squares}, and {\it filled circles} from HJG and this paper.} 
\label{lightcurve}
\end{figure}

\clearpage

\begin{figure}
\epsscale{1.0}
\plotone{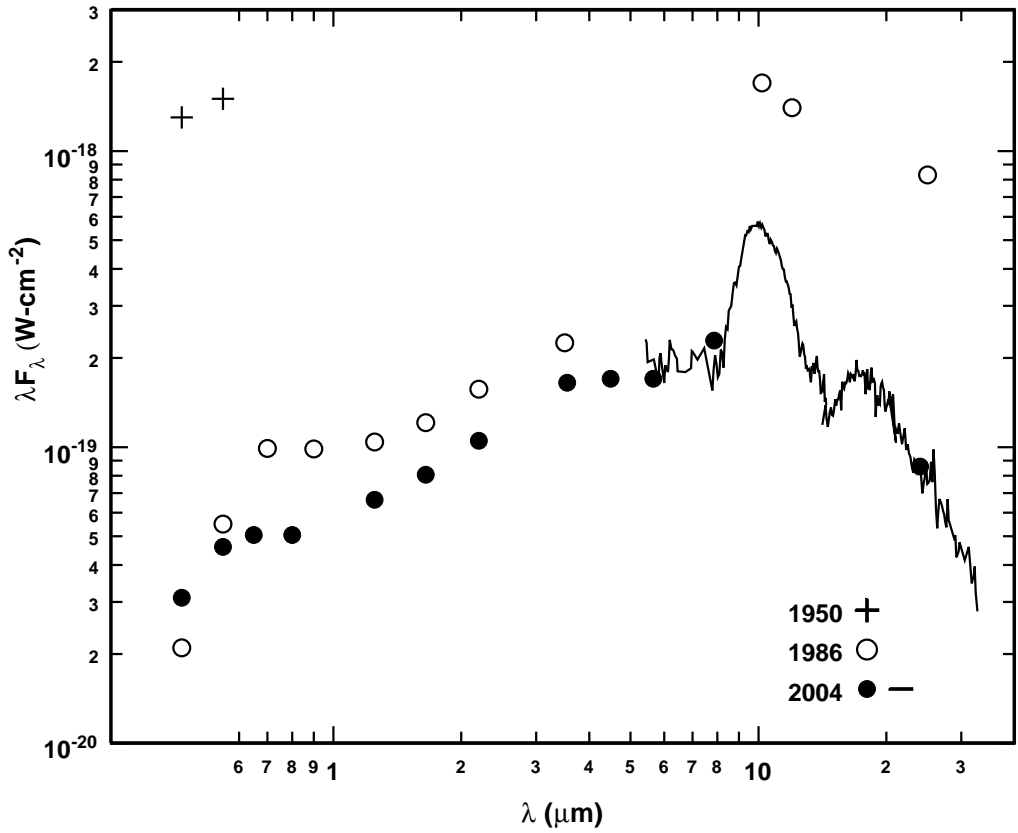}
\caption{Variable A's spectral energy distribution from 1986(HJG) and its current energy distribution from data in this paper. The IRS spectrum from 5 to 28{\micron} is also shown. The crosses show its apparent magnitudes at maximum light from Hubble \& Sandage(1953) transformed to standard B and V magnitudes on the Johnson system.}
\label{SED}
\end{figure} 

\begin{figure}
\epsscale{1.0}
\plotone{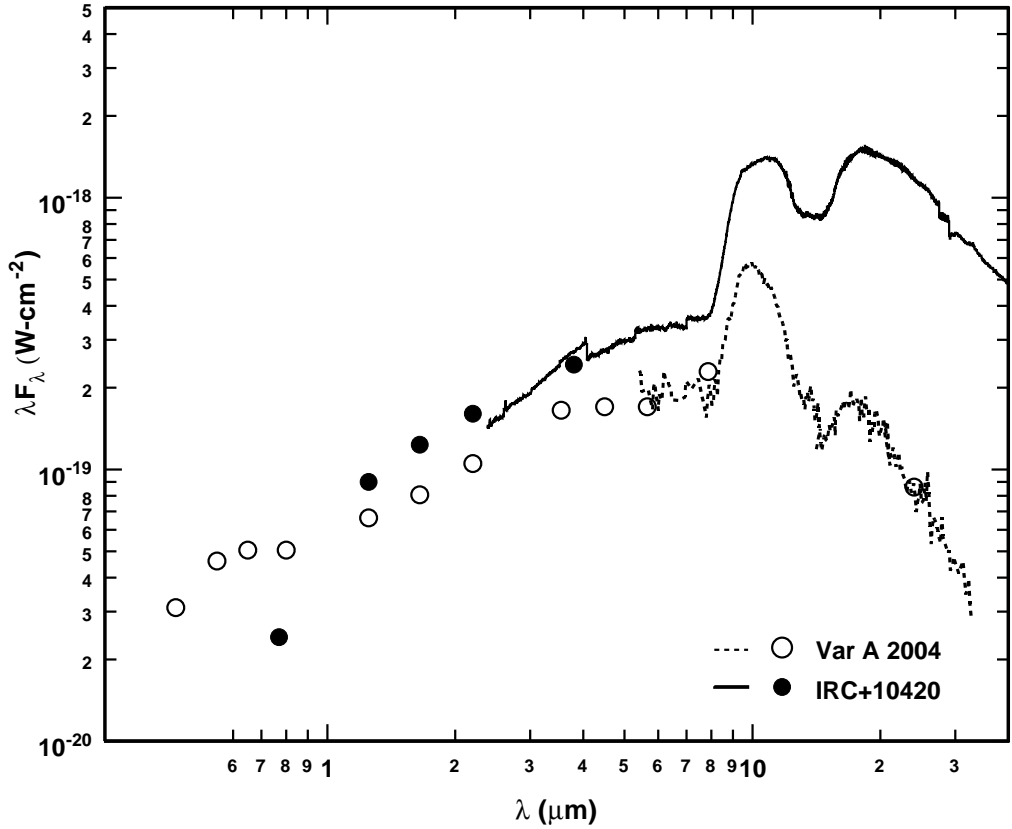}
\caption{A comparison of the spectral energy distributions of Variable A and IRC+10420 in our galaxy
shifted to the distance of M33. The spectrum of IRC~+10420 from 2 to 30 {\micron} is from
ISO data (Molster \etal 2002). The rapid drop at the shorter wavelengths in IRC~+10420 is due to high
interstellar extinction.}
\label{Comp}
\end{figure} 


\begin{deluxetable}{lcccc}
\tablecaption{Journal of New Observations}
\tablewidth{0pt}
\tablehead{
\colhead{Date(U.T.)}  &  \colhead{Grating/Filter}  &  \colhead{$\lambda_{0}$}  &
 \colhead{Spectral Resolution}   &  \colhead{Total Integration}
 }
 \startdata
 \sidehead{a. MMT Double Spectrograph}
 Nov. 18, 2003  &  600l(red)    &    6400{\AA}  &  4.7{\AA}   &   40$^{m}$\\
 Nov. 18, 2003  &  270l(red)    &    7000{\AA}  & 10.8{\AA}   &   60$^{m}$\\
 Nov. 19, 2003  &  300l(blue)   &    6000{\AA}  &  6.2{\AA}   &   45$^{m}$\\
 Sep. 23, 2004  &  300l(blue)   &    6000{\AA}  &  6.2{\AA}   &   80$^{m}$\\

 \sidehead{b. Polarimetry}
 Sep. 27, 2004 &  VRI           &    --   &   --    & 100$^{m}$,80$^{m}$,80$^{m}$
  \\

  \sidehead{c. Spitzer - IRAC}
  Jan. 9, 2004  &  3.6, 4.5, 5.8, 8.0{\micron}  &   --    &   --  &  31.2$^{s}$ ea
  ch\\
  Jul. 22, 2004 &         "                  &   --    &   --  &     "    \\
  Aug. 16, 2004 &         "                   &   --    &   --  &     "    \\
  Jan. 1, 2005  &         "                   &  --     &  --   &     "   \\

  \sidehead{d. Spitzer - IRS}
  Aug. 31, 2004 &  5 -- 15{\micron}  &  -- &  --  &  366$^{s}$   \\
       "        &  14 -- 28{\micron} &   --  &    --   &   731$^{s}$   \\

       \sidehead{e. Spitzer - MIPS}
       Dec. 29, 2003  & 24, 70, 160{\micron}  &   -- &     --  & 238$^{s}$, 43$^{s}$, 1
       9$^{s}$,              \\
       Feb. 3, 2005   &       "                &   -- &     --  &   "\\
       Sept. 5, 2005  &     "                &   --  &   --   &  " \\
       \enddata
       \end{deluxetable}

\begin{deluxetable}{lccccc}
\tablecaption{Summary of Photometric Data}
\tablewidth{0pt}
\tablehead{
\colhead{Bandpass} & \colhead{1986 (HJG)}  &   \colhead{1992}  &  \colhead{1997\tablenotemark{a}} & \colhead{2000 -- 01\tablenotemark{b}} & \colhead{2004 -- 05} \\
&   \colhead{mag.}  &  \colhead{mag.}  &   \colhead{mag.} & \colhead{mag.} & \colhead{mag.}
}
\startdata
U    &   --     & --    &  -- &  20.2, 20.1 $\pm$ 0.02 &  -- \\
B    &   20.4 $\pm$ 0.15   &  --   & -- & 19.9, 19.8 $\pm$ 0.01 & 19.9\tablenotemark{c}  \\
V    &   18.8 $\pm$ 0.10  &  --   & -- & 18.8, 18.8 $\pm$ 0.01 &  19.1\tablenotemark{c}, 19.1 $\pm$ 0.17\tablenotemark{d}  \\
R    &   17.7 $\pm$ 0.08   &  --   & -- &  18.2, 18.3 $\pm$ 0.01 & 18.5\tablenotemark{c}, 18.6 $\pm$ 0.06\tablenotemark{d}  \\
I(.8{\micron})  &  --   &  -- & -- & 17.7, 17.8 $\pm$ 0.01 & 18.1 $\pm$ 0.04\tablenotemark{d}\\
I(.9{\micron})  &  17.2 $\pm$ 0.10   & -- &  -- & -- & -- \\

\\

J    &   16.4$\pm$ 0.10   &   16.4 $\pm$ 0.13  &   16.9 $\pm$ 0.14  & -- & --  \\
H    &   15.5$\pm$ 0.10  &   15.3 $\pm$ 0.11  &    15.9 $\pm$ 0.16  & -- & --  \\
K    &   14.4$\pm$ 0.05  &   14.5 $\pm$ 0.10  &    14.7 $\pm$ 0.09  & -- & -- \\
L    &   12.6$\pm$ 0.10   &   --    &    --  & -- & 12.9 $\pm$ 0.05\tablenotemark{e}
\\

\\

4.5{\micron}  &   -- &   -- &  -- & -- &  12.1 $\pm$ 0.05\tablenotemark{e}   \\
5.7{\micron}  &   -- &   -- &  -- & -- &   11.4 $\pm$ 0.10\tablenotemark{e}    \\
8.0{\micron}  &   -- &   -- &  -- & -- &   10.1 $\pm$ 0.05\tablenotemark{e}    \\
10.2{\micron}  &  7.1 $\pm$ 0.20   &   -- &  -- & -- &  8.3\tablenotemark{f}   \\
24{\micron}    &  --  &   --  &  -- & -- &    7.5 $\pm$ 0.10     \\
70{\micron}    &  --  &   --  &  -- & -- &    4.3 \tablenotemark{g}   \\
160{\micron}    &  --  &   --  &  -- & -- &   -2.5 \tablenotemark{g}  \\
\enddata
\scriptsize
\tablenotetext{a}{2MASS, Dec., 1997}
\tablenotetext{b}{CCD photometry from Massey(2005) obtained Oct 2 \& 4, 2000 and Sep. 18, 2001, respectively.}
\tablenotetext{c}{From the flux-calibrated MMT spectrum}
\tablenotetext{d}{From the polarimetry frames}
\tablenotetext{e}{Spitzer -- IRAC. Here we give the mean of the four observations. The calibration uncertainties dominate the photometric errors of the individual observations; therefore the errors of the separate observations are the same. When the dispersion in the mean magnitude is less than the calibration error we quote the latter. In only one case, Ch 3 (5.7{\micron}), the error of the mean was greater, 0.10 mag.}
\tablenotetext{f}{From integration of the IRS spectrum over the N band filter.}
\tablenotetext{g}{The upper limit in magnitudes corresponding to 13.98 mJy and 1.5 Jy
 at 70{\micron} and 160{\micron}, respectively. The fluxes were converted to
  magnitudes using the Vega spectral model (Cohen 1996) and integrating over
      the bandpass.}
      \end{deluxetable}

\begin{deluxetable}{lcc}
\tablecolumns{3}
\tablecaption{The Strong Emission lines and Representative Absorption Lines in Variable A}
\tablewidth{0pt}
\tablehead{
\colhead{Line Id.} & \colhead{Hel. Vel.}  &   \colhead{W$_{\lambda}$}   \\
    &   \colhead{km s$^{-1}$}  &  \colhead{{\AA}}
    }
    \startdata
    \sidehead{Emission}
    H$\alpha$   &   -150.2\tablenotemark{a}     &    51.1\tablenotemark{b}, 52.1\tablenotemark{c}\\
    H$\beta$    &   -155.8\tablenotemark{d}     &     9.4\tablenotemark{b}, 10.9\tablenotemark{c}\\
    H$\gamma$   &   -147.6                      &     1.9 \\

    Ca~II $\lambda$8498  &   -130.7              &       4.4  \\
    Ca~II $\lambda$8542  &   -130.4             &        6.0  \\

    {[Ca~II]} $\lambda$7291 &   -149.9            &        1.9   \\
    {[Ca~II]} $\lambda$7324 &   -170.5            &        2.5    \\

    K~I $\lambda$7665     &   -96.0             &        1.9 \\
    K~I $\lambda$7699     &  -116.5             &        3.4\\

    \\
    \sidehead{Absorption}
    Ca~II H               &   -165.3            &         5.2\\
    Ca~II K               &   -170.2            &        10.0\\
    Fe~I $\lambda$4046   &   -174.1            &         1.4\\
    Fe~I $\lambda$4064   &   -141.0            &         0.4     \\
    Sr~II $\lambda$4077  &   -140.9            &         1.3   \\
    Ca~I  $\lambda$4226  &   -218.7            &         1.0 \\
    Fe~I $\lambda$4250   &   -195.0            &         0.7  \\
    Fe~I $\lambda$4271   &   -145.7            &         1.1   \\
Fe~I $\lambda$4325   &   -140.3           &          0.8  \\
Fe~I $\lambda$4383   &   -167.2           &          0.9 \\
Mg~I $\lambda$5167   &   -161.2           &          --     \\
Na~I~(D) $\lambda$5890 & -191.1           &          1.1 \\
\enddata
\scriptsize
\tablenotetext{a}{Thomson scattering wings at -975 and +882 km s$^{-1}$}
\tablenotetext{b}{without wings}
\tablenotetext{c}{with wings}
\tablenotetext{d}{Thomson scattering wings at -770 and +794 km s$^{-1}$}
\end{deluxetable}

\begin{deluxetable}{ccl}
\tablecolumns{3}
\tablecaption{The Stronger Uncertain and Unidentified Emission and  Absorption Lines}\tablewidth{0pt}
\tablehead{
\colhead{Measured $\lambda$}  &  \colhead{Predicted $\lambda$\tablenotemark{a}} & \colhead{Comment}\\
{\AA}    &    {\AA}      &
}
\startdata
\sidehead{Emission}
4359.5     &    4361.7  &   Probable blend of [Fe II] $\lambda$4359.3 and [OIII]  $\lambda$4363.2 \\
           &      &          from nearby nebulosity\\
	   4921.3     &   4823.8  &   very broad feature; a blend of several lines probably\\
	              &     & including Fe II  $\lambda$4923.9, [OIII]  $\lambda$4931.8, [FeII]
		       $\lambda$4924.5\\
		       5004.5     &   5007.0  &   [OIII] $\lambda$5006.8, possibly blended with [FeII]\\
		       5017.1     &   5019.6  &   Fe II $\lambda$5018.4, [FeII] $\lambda$5020.2 \\
		       5038.6     &   5041.1  &   possibly Si II $\lambda$5041.1\\
		       5079.4     &   5081.9  &    unidentified\\
		       5510.5     &   5513.3  &     very narrow\\
		       6397.3     &   6400.5  &     unidentified\\
		       6495.4     &   6498.6  &     unidentified\\
		       7247.5     &   7251.1  &     unidentified\\

		       \sidehead{Absorption}
		       3823.2     &  3825.5   &     unidentified\\
		       3855.7     &  3858.0   &     unidentified\\
		       5057.1     &  5060.1   &     unidentified\\
		       7236.1     &  7240.4   &     unidentified\\
		       7273.0     &  7277.4   &     unidentified\\
		       7337.2     &  7341.6   &     unidentified\\
		       \enddata
		       \scriptsize
		       \tablenotetext{a}{Respectively determined using the mean velocity of the hydrogen emission lines (-150 km s$^{-1}$) and the mean velocity of the absorption lines (-167.6
		       km s$^{-1}$).}
		       \end{deluxetable}


\begin{thebibliography}{}    

\bibitem[Allen (1963)]{Allen63} Allen, C. W. 1963, Astrophysical Quantities

\bibitem[Beardsley(1961)]{beardsley61} Beardsley, W.A. 1961, \apjs, 5, 381

\bibitem[Bidelman \& McKellar(1957)]{bidelman57} Bidelman, W.P. \& McKellar, A. 1957, \pasp, 69, 31

\bibitem[Cohen \etal (1996)]{Cohen96} Cohen, M., Witteborn, F. C., Carbon, D. F., Davies, J. K., Wooden, D. H., \&  Bregman, J, D. 1996, \aj, 112, 227 

\bibitem[de Jager(1998)]{dejager98} de Jager, C. 1998, \aapr, 8, 145 

\bibitem[Elitzur (1981)]{Elitzur81} Elitzur, M. 1981, in {\it Physical Processes in Red Giants}, ed. I. Iben \& A. Renzini (Reidel, Dordrecht), p. 363 

\bibitem[Gordon \etal (2005)]{KG05} Gordon, K. \etal 2005, \pasp, 117, 503

\bibitem[Gottleib \& Liller (1978)]{GL78} Gottleib, E. W. \& Liller, W. 1978, \apj, 225, 488

\bibitem[Hubble \& Sandage(1953)]{HS53} Hubble, E. \& Sandage, A. 1953, \apj, 118, 353 

\bibitem[Humphreys(1975)]{RMH75} Humphreys, R. M. 1975, \apj, 200, 426

\bibitem[Humphreys(1978)]{RMH78} Humphreys, R. M. 1978, \apj, 219, 445 

\bibitem[Humphreys \& Davidson(1979)]{HD79} Humphreys, R.M. \& Davidson, K. 1979,
 \apj, 232, 409 

\bibitem[Humphreys, Jones \& Gehrz(1987)]{HJG} Humphreys, R.M., Jones, T. J. \& Gehrz, R.D. (HJG) 
1987, \aj, 94, 315 

\bibitem[Humphreys \& Davidson(1994)]{humphreys94} Humphreys, R.M. \& Davidson, K. 1994,
 \pasp, 106, 1025

\bibitem[Humphreys \etal(1997)]{humphreys97} Humphreys, R.M., Smith, N., Davidson, K. 
 {\etal} ~1997, \aj, 114, 2778

\bibitem[Humphreys, Davidson \& Smith(2002)]{humphreys02} Humphreys, R.M., Davidson, K.,
 \& Smith, N. 2002, \aj, 124, 1026

\bibitem[Humphreys \etal(2005)]{humphreys05} Humphreys, R.M., Davidson, K., Ruch, G., \&
  Wallerstein, G. 2005, \aj, 129, 492


\bibitem[Johnson \& Jones(1991)]{JJ91} Johnson, J.~J., \&
Jones, T.~J.\ 1991, \aj, 101, 1735

\bibitem[Jones \etal(1993)]{TJJ93} Jones, T. J., Humphreys, R. M., Gehrz, R. D. {\etal} 1993, \apj, 411, 323

\bibitem[Kurtev, Corral, \& Georgiev (1999)]{Kurt99} Kurtev, R. G., Corral, L. G. \& Georgiev, L. 1999, \aap, 796

\bibitem[Landolt(1983)]{Landolt83} Landolt, A.~U.\ 1983, \aj, 88,
439

\bibitem[Little-Marenin \& Little(1990)]{LML90}
Little-Marenin, I.~R., \& Little, S.~J.\ 1990, \aj, 99, 1173


\bibitem[Lobel (1997)]{Lobel97} Lobel, A. 1997, Ph.D Thesis, Vrije Univ.

\bibitem[Lobel \etal(2003)]{lobel03} Lobel, A., Dupree, A.K., Stefanik, R.P. \etal~2003,
 \apj, 583, 923

\bibitem[Makovoz, Khan \& Moshir (2005)]{Mak05} Makovoz, D., Khan, I., \& Moshir, M. 2005, \pasp, 117, 274  

\bibitem[Massey \etal(2002)]{Massey02} Massey, P., Hodge, P. W., Holmes, S., Jacoby, J., King, N. L., Olsen, K., Smith, C., \& Saha, A.  2002, \baas, 34, 1272

\bibitem[Massey (2005)]{PM05} Massey, P. 2005, private communication


\bibitem[Molster et al.(2002)]{Molster02} Molster, F.~J., Waters, L.~B.~F.~M., Tielens, A.~G.~G.~M., \& Barlow, M.~J.\ 2002, \aap, 382, 184


\bibitem[Newton (1980)]{New80} Newton, K. 1980, \mnras, 190, 689

\bibitem[Nuth \& Hecht(1990)]{Nuth90} Nuth, J.~A., \& Hecht,
J.~H.\ 1990, \apss, 163, 79

\bibitem[Reach \etal (2005)]{Rea05} Reach \etal 2005, \pasp, 117, 978 

\bibitem[Rice \etal(1990)]{Rice90} Rice, W., Boulanger, F., Viallefond, F., Soifer, B. T. \& Freedman, W. L. 1990, \apj, 358, 418  

\bibitem[Schmidt et al.(1992)]{Schmidt92} Schmidt, G.~D., Elston, 
R., \& Lupie, O.~L.\ 1992, \aj, 104, 1563 

\bibitem[Schuster, Humphreys \& Marengo (2005)]{Schuster05} Schuster, M, Humphreys, R. M. \& Marengo, M. 2006, \aj, in press 

\bibitem[Smith \etal(2002)]{Smith02} Smith, N., Davidson, K., Gull, T.R., Ishibashi, K. \& Hillier, D. J. 2003, \apj, 586, 432

\bibitem[Wallerstein 1958]{Wall58}Wallerstein, G. 1958, \pasp, 70, 479

\end{thebibliography}
\end{document}